\documentclass[prl,aps,floats,amsmath,amssymb,twocolumn,showpacs,superscriptaddress,
preprintnumbers,nofootinbib]{revtex4}

\usepackage{graphicx}
\usepackage{epstopdf}

\newcommand{\be}{\begin{equation}}
\newcommand{\ee}{\end{equation}}
\def\ltorder{\mathrel{\raise.3ex\hbox{$<$}\mkern-14mu
 \lower0.6ex\hbox{$\sim$}}}
\def\gsim{\mathrel{\rlap{\lower4pt\hbox{\hskip1pt$\sim$}}
    \raise1pt\hbox{$>$}}}                

\begin{document}

\preprint{
\vbox{
\hbox{BNL-HET-07/19, Edinburgh 2007/38, MKPH-T-07-16, RBRC-699, SHEP-0743} 
}}

\title{$K_{l3}$ semileptonic form factor from $2+1$ flavor lattice
  QCD}

\author{P.~A.~Boyle}
\affiliation{School of Physics, University of Edinburgh,
  Edinburgh EH9 3JZ, UK}
\author{A.~J\"uttner}
\affiliation{Institut f\"ur Kernphysik,
Johannes Gutenberg-Universit\"at Mainz,
D-55099 Mainz, Germany}
\author{R.~D.~Kenway}
\affiliation{School of Physics, University of Edinburgh,
  Edinburgh EH9 3JZ, UK}
\author{C.~T.~Sachrajda}
\affiliation{School of Physics and Astronomy, University of
  Southampton,  Southampton, SO17 1BJ, UK}
\author{S.~Sasaki}
\affiliation{RIKEN-BNL Research Center, Brookhaven National
  Laboratory,  Upton, NY 11973, USA}
\affiliation{Department of Physics, University of Tokyo, Tokyo
  113-0033,  Japan}
\author{A.~Soni}
\affiliation{Physics Department, Brookhaven National Laboratory,
  Upton, NY 11973, USA}
\author{R.~J.~Tweedie}
\affiliation{School of Physics, University of Edinburgh,
  Edinburgh EH9 3JZ, UK}
\author{J.~M.~Zanotti}
\affiliation{School of Physics, University of Edinburgh,
  Edinburgh EH9 3JZ, UK}
\collaboration{RBC+UKQCD Collaborations} \noaffiliation
\begin{abstract}
  We present the first results for the $K_{l3}$ form factor from
  simulations with $2+1$ flavors of dynamical domain wall quarks.
  Combining our result, namely $f_+(0)=0.964(5)$, with the latest
  experimental results for $K_{l3}$ decays leads to
  $|V_{us}|=0.2249(14)$, reducing the uncertaintity in this important
  parameter.
  For the $O(p^6)$ term in the chiral expansion we obtain $\Delta
  f=-0.013(5)$.
\end{abstract}


\maketitle

The increasing precision with which the unitarity of the CKM matrix
\cite{Cabibbo:1963yz} can be tested is an important tool for exploring
the limits of the Standard Model.
One such unitarity relation is
\begin{equation}
|V_{ud}|^2+|V_{us}|^2+|V_{ub}|^2=1\,,
\end{equation}
whose uncertainty is dominated by the precision of $|V_{us}|$.
In order to obtain $|V_{us}|$ from experimental measurements of the rate
for an $s\to u$ decay process, it is necessary to quantify the
corresponding non-perturbative QCD effects.
In this paper we present first lattice results from simulations with
$2+1$ flavors of domain wall quarks, which respect chiral and flavor
symmetries to high accuracy, for the evaluation of the
form factor $f_+(0)$ necessary to determine $|V_{us}|$ from
$K\to\pi\ell\nu_\ell$ ($K_{\ell 3}$) semileptonic decays.
Precise knowledge of $f_+(0)$ is crucial also for deducing $|V_{td}|$
from a measurement of $K \to \pi^0 \nu \bar \nu$.

Our determination of $f_+(0)$ includes estimates of all systematic
errors (chiral and $q^2$ extrapolations, discretization and finite
volume effects), and reduces the combined theoretical and
experimental error in $|V_{us}|$ from the PDG(2006) result of
0.2257(21) to
\footnote{For a recent review of lattice determinations of $f_K$ (necessary for
the determination of $|V_{us}|$ from $K_{\ell 2}$ leptonic decays) and
previous computations of $f_+(0)$ 
see \cite{juettnerlatt2007}.}
\be
|V_{us}| = 0.2249(14)\,.
\ee

The combination $|V_{us}f_+(0)|$ can be obtained from the experimental
rate for $K_{\ell 3}$ decays
\begin{eqnarray}
\Gamma_{K\to\pi l\nu} &=& C_K^2\frac{G_F^2 m_K^5}{192\pi^3} I\,
S_{EW}   \\
&\times& \left[1+2\Delta_{SU(2)} +2\Delta_{EM}\right]
|V_{us}|^2|f_+(0)|^2\, , \nonumber
\end{eqnarray}
where $I$ is the phase space integral which can be evaluated from the
shape of the experimental form factor, and
$\Delta_{SU(2)},\,S_{EW},\,\Delta_{EM}$ contain the isospin breaking,
short distance electroweak and long distance electromagnetic
corrections, respectively.
$f_+(0)$ is the form factor defined from the $K\to \pi$ matrix element
of the weak vector current, $V_\mu=\bar{s}\gamma_\mu u$, evaluated at
zero momentum transfer
\be
\langle \pi(p^\prime) \big | V_\mu \big | K(p)\rangle = (p_\mu +
p_\mu^\prime) f_+(q^2) + (p_\mu - p_\mu^\prime) 
f_-(q^2)\,,
\label{eq:ME}
\ee
where $q^2=(p-p^\prime)^2$.
PDG(2006) quotes \cite{blucher06}\footnote{A more recent analysis
  finds $|V_{us}f_+(0)|=0.21673(46)$~\cite{Moulson:2007fs}.}
\be
|V_{us}f_+(0)|=0.2169(9)\ ,
\ee
hence in order to obtain $|V_{us}|$ at a precision commensurate with
current experiments, we need to determine $f_+(0)$ with an error of
less than 1\%.

In chiral perturbation theory (ChPT), $f_+(0)$ is expanded in terms of
the light pseudoscalar meson masses
\be
f_+(0) = 1 + f_2 + f_4 + \ldots,\quad (f_n={\cal
  O}(m^n_{\pi,\,K,\,\eta}))\ .
\label{eq:chiralexp}
\ee
Current conservation ensures that in the $SU(3)_\text{flavor}$ limit
$f_+(0)=1$, hence $f_2$ and $f_4$ are \textit{small}.
Additionally, as a result of the Ademollo-Gatto Theorem
\cite{Ademollo:1964sr}, which states that $f_2$ receives no
contribution from local operators appearing in the effective theory,
$f_2$ is determined unambiguously in terms of $m_\pi$, $m_K$ and
$f_\pi$, and takes the value $f_2=-0.023$ at the physical values of
the meson masses~\cite{Leutwyler:1984je}.
Our task is now reduced to one of finding
\be
\Delta f = f_+(0) - (1 + f_2) \ .
\label{eq:deltaf}
\ee
Until recently, the canonical estimate of $\Delta f=-0.016(8)$ was due
to Leutwyler \& Roos (LR) \cite{Leutwyler:1984je}, whereas more recent
ChPT based phenomenological analyses favor a value consistent with
zero (see Table~\ref{table:lattresults}).
These determinations, however, require model input; the 50\% error in
the LR result, for example, was estimated within the context of a
simple quark model.
Hence a model independent determination of $\Delta f$ with a reliable
error estimate is necessary.
We compile recent lattice and phenomenological results in
Table~\ref{table:lattresults}.
Our lattice calculation has been discussed in preliminary form in
\cite{Antonio:2007mh} and we now finalize our results with the
inclusion of the complete set of data and
a careful estimate of all systematic errors.

\begin{table}
\begin{center}
  \caption{Summary of ChPT and lattice results. ${}^\dagger$
    Results in conference proceedings only. $^*$ Used slope of
    experimental form factor as input. $\ddag$~Information not
    provided.  
}
  \label{table:lattresults}
  \begin{tabular}{c@{\hspace{1mm}}|@{\hspace{2mm}}l@{\hspace{1mm}}lcc@{\hspace{1mm}}c}
\hline
      Ref. & \hspace{2mm}$f_+(0)$ & \hspace{4mm}$\Delta f$ & $m_\pi$ [GeV] &
      $a$ [fm] & $N_f$ 
  \\ \hline
 \cite{Leutwyler:1984je}\phantom{$^\dagger$$^*$}  & 0.961(8)      & -0.016(8)  &&&\\
 \cite{Bijnens:2003uy}\phantom{$^\dagger$$^*$}    & 0.978(10)     & +0.001(10)  &&&\\
 \cite{Cirigliano:2005xn}\phantom{$^\dagger$$^*$} & 0.984(12)     & +0.007(12) &&&\\
 \cite{Jamin:2004re}\phantom{$^\dagger$$^*$}      & 0.974(11)     & -0.003(11) &&&\\
\hline
 \cite{Becirevic:2004ya}\phantom{$^\dagger$$^*$}&0.960(5)(7)&-0.017(5)(7)&$\gsim 0.5$ & 0.07  & 0 \\
 \cite{Dawson:2006qc}\phantom{$^\dagger$$^*$} & 0.968(9)(6)&-0.009(9)(6)&$\gsim 0.49$& 0.12  & 2 \\
 \cite{Okamoto:2004df}$^\dagger$$^*$& 0.962(6)(9)&-0.015(6)(9)&$\ddag$&$\ddag$& 2+1 \\
 \cite{Tsutsui:2005cj}$^\dagger$\phantom{$^*$}& 0.967(6)&-0.010(6)&$\gsim 0.55$& 0.09  & 2 \\
 \cite{Brommel:2007wn}$^\dagger$\phantom{$^*$}& 0.965(2)&-0.012(2)&$\gsim 0.5$ & 0.08  & 2 \\
 This work               & 0.964(5)   & -0.013(5)  & $\gsim 0.33$ & 0.114 & 2+1 \\
\hline
  \end{tabular}
\end{center}
\end{table}

We simulate with $N_f=2+1$ dynamical flavors generated with the
Iwasaki gauge action \cite{Iwasaki:1985we} at $\beta=2.13$, which
corresponds to an inverse lattice spacing $a^{-1}=1.73(3)\,\text{GeV}$
($a=0.114(2)\,\text{fm}$)~\cite{Allton:2007hx,24cubed}, and the domain
wall fermion action \cite{Kaplan:1992bt} with a residual mass of
$am_\text{res}= 0.00315(2)$ \cite{Allton:2007hx,24cubed}.
The simulated strange quark mass, $am_s=0.04$, is close to its
physical value \cite{24cubed}, and we choose four values for the light
quark masses, $am_{ud}$, which correspond to pion masses as light as
329 MeV \cite{Allton:2007hx,24cubed}.
The calculations are performed on two volumes, $16^3$
($(1.83)^3\,\text{fm}^3$) and $24^3$ ($(2.74)^3\,\text{fm}^3$), at
each quark mass, except the lightest mass which is only simulated on
the larger volume.
Simulation details are summarized in Table~\ref{table:parameters} and
more details can be found in \cite{Allton:2007hx,24cubed}.

We start by rewriting the vector form factors given in (\ref{eq:ME})
to define the scalar form factor
\be
f_0(q^2) = f_+(q^2) + \frac{q^2}{m_K^2 - m_\pi^2}f_-(q^2)\ ,
\ee
which can be obtained on the lattice at $q^2_\text{max}=(m_K -
m_\pi)^2$ with high statistical accuracy
\cite{Becirevic:2004ya,Hashimoto:1999yp}.
In Table~\ref{table:q2max} we present our results for
$f_0(q^2_\text{max})$ for each of the simulated quark masses and
volumes.

\begin{table}[t]
\begin{center}
  \caption{Simulation parameters: bare light quark mass ($am_{ud}$),
    pion ($m_{\pi}$) and kaon ($m_K$) masses for both volumes.}
  \label{table:parameters}
  \begin{tabular}{l@{\hspace{2mm}}|@{\hspace{2mm}}l@{\hspace{4mm}}l@{\hspace
{2mm}}|@{\hspace{2mm}}l@{\hspace{4mm}}l}
\hline
\multicolumn{1}{c}{}&\multicolumn{2}{c}{$16^3\times 32$} &
\multicolumn{2}{c}{$24^3\times 64$}
  \\ \hline
      $am_{ud}$ & $m_{\pi}$~[GeV] & $m_K$~[GeV] &
      $m_{\pi}$~[GeV] & $m_K$~[GeV]
  \\ \hline
  0.03  & 0.674(11)   & 0.723(12)   & 0.671(11) & 0.719(12) \\
  0.02  & 0.557(9)    & 0.666(11)   & 0.556(9)  & 0.663(11) \\
  0.01  & 0.428(7)    & 0.614(10)   & 0.416(7)  & 0.604(10) \\
  0.005 &\hspace{4mm}-&\hspace{4mm}-& 0.329(5)  & 0.575(9) \\
\hline
  \end{tabular}
\end{center}
\end{table}

For each quark mass, in addition to evaluating $f_0(q^2)$ at
$q^2=q^2_\text{max}$, we determine the form factor at several negative
values of $q^2$, allowing us to interpolate to $q^2=0$.
Specifically, in the notation of (\ref{eq:ME}), we evaluate the form
factor with $|\vec{p}\,'|=0$, $|\vec{p}\,|=p_L$ or
$|\vec{p}\,|=\sqrt{2}\,p_L$ where $p_L=2\pi/L$ and $L$ is the spatial
extent of the lattice, and also with $|\vec{p}\,|=0$,
$|\vec{p}\,'|=p_L$ or $|\vec{p}\,'|=\sqrt{2}\,p_L$\,.
To obtain the $f_0(q^2)$ we use standard ratio techniques
\cite{Hashimoto:1999yp,Becirevic:2004ya,Dawson:2006qc}, which do not
require normalization of the vector current.

\begin{table}[b]
\begin{center}
  \caption{Results for $f_0(q^2_\text{max})$ where
    $q^2_\text{max}=(m_K - m_\pi)^2$.}
  \label{table:q2max}
  \begin{tabular}{l@{\hspace{1mm}}|@{\hspace{1mm}}l@{\hspace{2mm}}l@{\hspace
{1mm}}|@{\hspace{1mm}}l@{\hspace{2mm}}l}
\hline
\multicolumn{1}{c}{}&\multicolumn{2}{c}{$16^3\times 32$} &
\multicolumn{2}{c}{$24^3\times 64$}
  \\ \hline
$am_{ud}$ &$q^2_\text{max}$~[GeV$^2$] & $f_0(q^2_\text{max})$
&$q^2_\text{max}$~[GeV$^2$]  & $f_0(q^2_\text{max})$  \\ \hline
  0.03  & 0.00233(4)  & 1.00035(3)  & 0.00235(4)   & 1.00029(6)   \\
  0.02  & 0.01178(24) & 1.00241(19) & 0.01152(20)  & 1.00192(34)  \\
  0.01  & 0.03475(66) & 1.01436(81) & 0.03524(62)  & 1.00887(89)  \\
  0.005 &\hspace{7mm}-&\hspace{7mm}-& 0.06070(107) & 1.02143(132) \\
  \hline
  \end{tabular}
\end{center}
\end{table}

In order to gain the maximum amount of information from limited data,
we perform a simultaneous fit to both the $q^2$ and quark mass
dependencies using the ansatz
\begin{eqnarray}
\lefteqn{f_0(q^2,m_\pi^2,m_K^2) = }\nonumber\\
&&\frac{1+f_2+(m_K^2-m_\pi^2)^2(A_0 + A_1(m_K^2+m_\pi^2))}
{1-q^2/(M_0+M_1(m_K^2+m_\pi^2))^2}\ ,
\label{eq:global}
\end{eqnarray}
with four fit parameters $A_0,\,A_1,\,M_0,\,M_1$, and where $f_2$ is
the NLO term appearing in the chiral expansion (\ref{eq:chiralexp}),
evaluated by inserting the lattice results for $m_\pi,\,m_K$ and the
physical value for $f_\pi$ (132~MeV) into the expression appearing in
ChPT \cite{Leutwyler:1984je} at each quark mass~\footnote{Here we note
  that by using $f_\pi$ in $f_2$, we are following convention
  \cite{Leutwyler:1984je}. The true SU(3) LEC $f_0$, however, is
  likely to be somewhat smaller, resulting in a larger and more
  dominant contribution coming from the corresponding $f_2$, and hence
  a more apparent convergence of the ChPT.  Our lattice calculation
  determines $1-f_+$ directly and will only differ slightly in the
  extrapolation ansatz.}.

The expression (\ref{eq:global}) is well motivated since we know from
the Ademollo-Gatto Theorem that to leading order $\Delta f \propto
(m_s - m_{ud})^2$, hence we expect
\be 
f_0(0) = 1 + f_2 + (m_K^2-m_\pi^2)^2 (A_0 +
A_1(m_K^2+m_\pi^2))\, ,
\label{eq:chiral1}
\ee
which incorporates the correct $SU(3)_\text{flavor}$ limit,
$f_+(0)=1$, to be a good phenomenological ansatz for the mass
dependence of $f_0(0)=f_+(0)$.
This motivates the numerator in (\ref{eq:global}), while the
denominator comes from simply including a quark mass dependence into
the standard pole dominance form
\be
f_0(q^2) = f_0(0)/(1-q^2/M^2)\ ,
\label{eq:monopole}
\ee
where $M$ is a pole mass, which has been shown to describe the
$q^2$-dependence of lattice results of $f_0(q^2)$ very well
\cite{Dawson:2006qc,Becirevic:2004ya}.

The traditional approach of sequentially interpolating in $q^2$
(\ref{eq:monopole}) followed by chiral extrapolation of $f_+(0)$
(\ref{eq:chiral1}) should agree with our simultaneous fit
(\ref{eq:global}).
Fitting the $24^3$ data only yields excellent agreement (shown in
the final two rows of Table~\ref{table:fzero}), with a reduced error
evident in the simultaneous fit, which we therefore take as our best
result.
For the $16^3$ data the pole fits generally have a poor $\chi^2$/dof.
We also find that the simultaneous and sequential fits to the $q^2$
and mass dependence for the $16^3$ data differ at $1.2\sigma$.
Consequently we only use the $16^3$ data to check that the
finite-volume effects are small. 
Tables~\ref{table:q2max} and \ref{table:fzero} demonstrate that this
is the case.

\begin{figure}[tb]
\includegraphics[width=8cm]{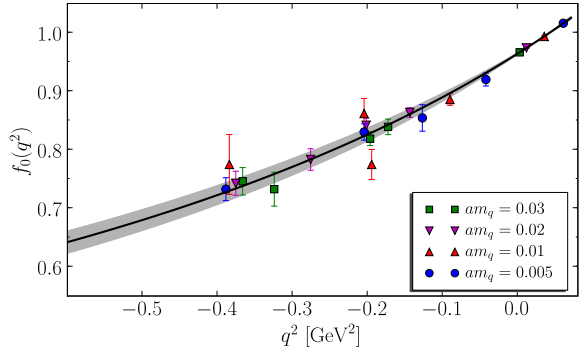}
\vspace*{-8mm}
\caption{Scalar form factor, $f_0(q^2)$, together with the simultaneous
  fit of (\ref{eq:global}) as described in the text.  }
\label{fig:global}
\end{figure}

\begin{table}[b]
\begin{center}
  \caption{Results for $f_+(0)$ using pole dominance
    (\ref{eq:monopole}) and quadratic (\ref{eq:quadratic}) 
    fits to each data set, together with the chiral extrapolations
    using (\ref{eq:chiral1}) with the $24^3\times 64$ data
    only. The final row gives the results for simultaneous 
$q^2$ and quark mass 
fits ((\ref{eq:global}) and (\ref{eq:globquad})) using the
    same data sets.}
  \label{table:fzero}
  \begin{tabular}{l@{\hspace{2.5mm}}|@{\hspace{2.5mm}}l@{\hspace{5mm}}r@{\hspace
{2.5mm}}|@{\hspace{2.5mm}}l@{\hspace{5mm}}r}
\hline
&\multicolumn{2}{c}{Pole} &
\multicolumn{2}{c}{Quadratic}
  \\ \hline
     $am_{ud}$ & $f_+(0)$ & $\chi^2$/dof &
      $f_+(0)$ & $\chi^2$/dof 
  \\ \hline
\multicolumn{5}{c}{$16^3\times 32$} \\
\hline
  0.03  & 0.99925(8) & 5.0/3  & 0.99938(12)& 4.2/2 \\
  0.02  & 0.9951(6)  & 13.5/3 & 0.9959(9)  & 13.0/2 \\
  0.01  & 0.9889(26) & 13.9/3 & 0.9866(33) & 10.9/2 \\
\hline
\multicolumn{5}{c}{$24^3\times 64$} \\
\hline
  0.03  & 0.9991(2)  & 2.1/3  & 0.9990(2)  & 1.5/2 \\
  0.02  & 0.9960(7)  & 2.3/3  & 0.9962(9)  & 1.9/2 \\
  0.01  & 0.9841(29) & 10.4/3 & 0.9806(39) & 7.7/2 \\
  0.005 & 0.9774(35) & 4.0/3  & 0.9749(59) & 2.7/2 \\
\hline
chiral   & 0.9644(39) & 3.4/2   & 0.9622(61) & 5.1/2 \\
sim. fit & 0.9644(33) & 28.7/16 & 0.9610(43) & 26.4/14 \\
\hline
  \end{tabular}
\end{center}
\end{table}

We present the results from a fit to the $24^3\times 64$ data sets
using (\ref{eq:global}) in Fig.~\ref{fig:global}.
Here the curve shows the fit function at the physical meson masses,
while the difference $f_0(q^2,m_\pi^{\rm latt},m_K^{\rm latt})$ $-$
$f_0(q^2,m_\pi^{\rm phys},m_K^{\rm phys})$ has been subtracted from our
raw data points and the small scatter is indicative of the quality of
our fit.

\begin{figure}[tb]
\includegraphics[width=8cm]{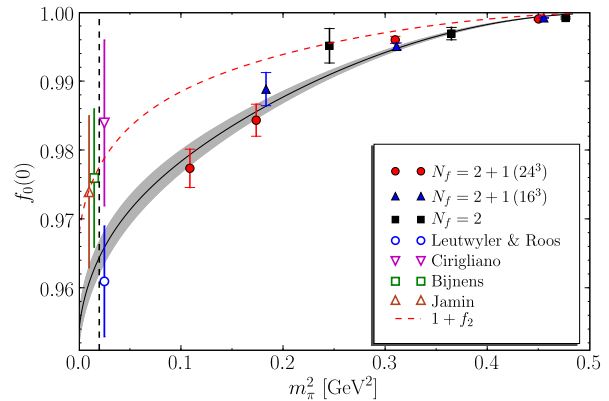}
\vspace*{-5mm}
\caption{Scalar form factor, $f_0(0)$,
  together with the simultaneous fit (solid line) on the $24^3$ data
  (red circles) using (\ref{eq:global}).}
\label{fig:global-chiral}
\end{figure}

The quark mass dependence of (\ref{eq:global}) is presented in
Fig.~\ref{fig:global-chiral}. 
The solid line represents the fit function evaluated at $q^2=0$,
plotted as a function of $m_\pi^2$, while the dashed line is the
contribution coming from the ${\cal O}(p^4)$ terms in the chiral
expansion, $1+f_2$.
Our results clearly indicate a sizeable, negative value for $\Delta f
=-0.013(3)$, in contrast to the recent ChPT based results
of~\cite{Bijnens:2003uy,Cirigliano:2005xn,Jamin:2004re}.
In Fig.~\ref{fig:global-chiral} we also overlay the results given in
Table~\ref{table:fzero} for $f_0(0)$ obtained from individual pole
fits on each of our ensembles and earlier $N_f=2$ results
\cite{Dawson:2006qc}.

So far, we have assumed a pole dominance behavior (\ref{eq:global})
in our lattice data, whose $q^2$ dependence differs marginally at NLO
from an expression obtained in ChPT \cite{Bijnens:2003uy}.
In order to estimate the systematic error due to this choice, we also
present in Table~\ref{table:fzero} results for fits to our data using
a quadratic ansatz
\be
f_0(q^2) = a_0 + a_1 q^2 + a_2 q^4\ ,
\label{eq:quadratic}
\ee
together with a chiral extrapolation using (\ref{eq:chiral1}).
A simultaneous fit similar to (\ref{eq:global}) is possible via
\begin{eqnarray}
\lefteqn{f_0(q^2,m_\pi^2,m_K^2) = }\nonumber\\
&&1+f_2+(m_K^2-m_\pi^2)^2(A_0 + A_1 + A_2(m_K^2+m_\pi^2)) +\nonumber\\
&&(A_3 +(2A_0+A_1)(m_K^2+m_\pi^2))\,q^2 +\nonumber\\
&&(A_4-A_0+A_5(m_K^2+m_\pi^2))\,q^4\ .
\label{eq:globquad}
\end{eqnarray}
The form of this ansatz is motivated by the expression obtained in
ChPT \cite{Bijnens:2003uy}.
We quote the result from a fit to the $24^3\times 64$ data using
(\ref{eq:globquad}) in the last row of Table~\ref{table:fzero}, where
we find that the results of the two fits, (\ref{eq:global}) and
(\ref{eq:globquad}), agree within statistical precision and we take
the difference (0.0034) as an estimate of the systematic error in
choosing (\ref{eq:global}) as our preferred ansatz.

Recently, an alternative parametrization, obtained by using
analyticity and crossing symmetry, has been proposed
\cite{Hill:2006bq}.
We find that fitting our data using this ansatz leads to results that
lie within the systematic uncertainty of 0.0034 discussed above.

Our simulations are performed with a strange quark mass
($am_s+am_\text{res}\simeq 0.043$) which is heavier than the physical
mass ($am_s+am_\text{res}\simeq 0.037$).
Both (\ref{eq:global}) and (\ref{eq:globquad}) are modelled according
to ChPT and this mass difference is corrected when we insert the
physical kaon mass to obtain our final result.
This correction is accurate in as much as our extrapolation model
describes our data, and any error introduced is included in our
estimate of the systematic error.
Future simulations will include a second valence strange quark
mass to decrease the reliance on our fit model.

\begin{figure}[!t]
\includegraphics[angle=-90,width=7.3cm]{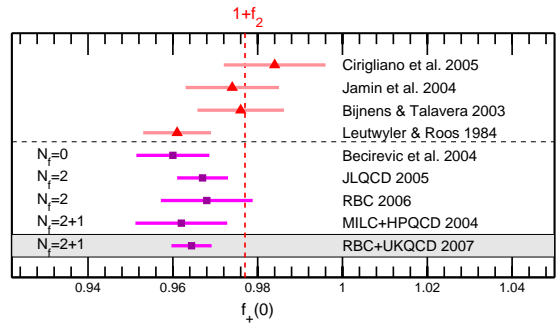}
\caption{Comparision with other determinations of $f_+(0)$.}
\label{fig:fzero}
\end{figure}

Finally, since we simulate at a single lattice spacing, we are
unable to extrapolate to the continuum limit.
However, leading lattice artefacts with domain wall fermions are of 
$O(a^2 \Lambda^2_{QCD})$; assuming $\Lambda_{QCD} \sim 300$~MeV we
estimate these to be no larger than $\approx 4\%$ (of $1-f_+$).
A comparison of the pion and kaon decay constants obtained from our
simulations with their physical values provides a test for the
reliablity of our result.
After including the effects to NLO due to chiral logs, we find $f_\pi$
and $f_K$ about 4\% below experiment~\cite{24cubed}, which is
consistent with our estimated scaling error.
We will explicitly check this for $K_{\ell 3}$ decays on our new
ensemble which is being generated on a finer lattice.
Note that our current uncertainty is dominated by statistics and the
chiral and $q^2$ extrapolations and not by the discretization error.
Hence our final result is
\be
f_+(0)=0.9644(33)(34)(14)\ ,
\ee
where the first error is statistical, and the second and third are
estimates of the systematic errors due to our choice of
parametrization (\ref{eq:global}) and lattice artefacts, respectively.
To put this result in context, we compare our value with other
determinations of $f_+(0)$ in Fig.~\ref{fig:fzero}.
We see that our result agrees very well with the Leutwyler-Roos
value~\cite{Leutwyler:1984je} and earlier lattice calculations
\cite{Becirevic:2004ya,Okamoto:2004df,Tsutsui:2005cj,Dawson:2006qc}.
In particular, we note that our findings prefer a sizeable, negative
value for $\Delta f=-0.0129(33)(34)(14)$, in contrast to recent ChPT
based phenomenological
results~\cite{Bijnens:2003uy,Cirigliano:2005xn,Jamin:2004re}.

Using $|V_{us}f_+(0)|=0.2169(9)$ from PDG(2006)
\cite{blucher06}\footnote{Using 
the result 
from \cite{Moulson:2007fs} gives $|V_{us}| =
0.2247(5)(11)$.}
\be
|V_{us}| = 0.2249(9)_\text{exp}(11)_{f_+(0)}\ ,
\ee
and combined with $|V_{ud}|=0.97377(27)$~\cite{blucher06} we find 
\be
|V_{ud}|^2 + |V_{us}|^2 + |V_{ub}|^2 = 1-\delta,\quad
\delta=0.0012(8)\,,
\ee
compared with the PDG(2006)~\cite{blucher06} result,
$\delta=0.0008(10)$.
Further reduction in the lattice error is imperative.
Our $q^2$ interpolation systematic is removable in principle
\cite{Boyle:2007wg} and we are in the process of addressing both this
and discretization systematics with a new set of simulations.

We thank D.\,Chen, N.\,Christ, M.\,Clark, S.\,Cohen, C.\,Cristian,
Z.\,Dong, A.\,Gara, A.\,Jackson, C.\,Jung,
C.\,Kim, L.\,Levkova, X.\,Liao, G.\,Liu, R.\,Mawhinney,
S.\,Ohta, K.\,Petrov, T.\,Wettig and A.\,Yamaguchi for developing with
us the QCDOC machine and its software.  This development and
resulting computer equipment were funded by
U.S.\ DOE grant DE-FG02-92ER40699, PPARC JIF grant
PPA/J/S/1998/00756 and RIKEN.  This work was supported by DOE
grants DE-FG02-92ER40699 and DE-AC02-98CH10886, PPARC grants
PPA/G/O/2002/00465, PP/D000238/1, PP/C504386/1, PPA/G/S/2002/00467 and
PP/D000211/1 and JSPS grant (19540265).
We thank BNL, EPCC, RIKEN, and the U.S.\ DOE for supporting the
essential computing facilities
We also thank D.\,Antonio, C.\,Dawson, T.\,Izubuchi, T.\,Kaneko,
C.\,Maynard and B.\, Pendleton for assistance.

\end{document}